\documentclass[prl,twocolumn,groupaddress]{revtex4}
\usepackage{graphicx}
\usepackage{epsfig,amsmath,amssymb,graphics,color,calc}
\newcommand{\be}{\begin{equation}}
\newcommand{\ee}{\end{equation}}
\newcommand{\ba}{\begin{eqnarray}}
\newcommand{\ea}{\end{eqnarray}}

\begin{document} 
\title{Dynamic heterogeneity in amorphous materials}

\author{Ludovic Berthier}
\affiliation{Laboratoire Charles Coulomb, 
UMR 5221 CNRS and Universit\'e Montpellier 2, Montpellier, France}

\date{\today}

\begin{abstract}

Amorphous solids are mechanically rigid while possessing a disordered
structure similar to that of dense liquids. Recent research indicates
that dynamical heterogeneity, spatio-temporal fluctuations in
local dynamical behavior, might help understanding 
the statistical mechanics of glassy states.

\end{abstract}

\maketitle

\section{The puzzle posed by amorphous materials}

\begin{figure*}
\psfig{file=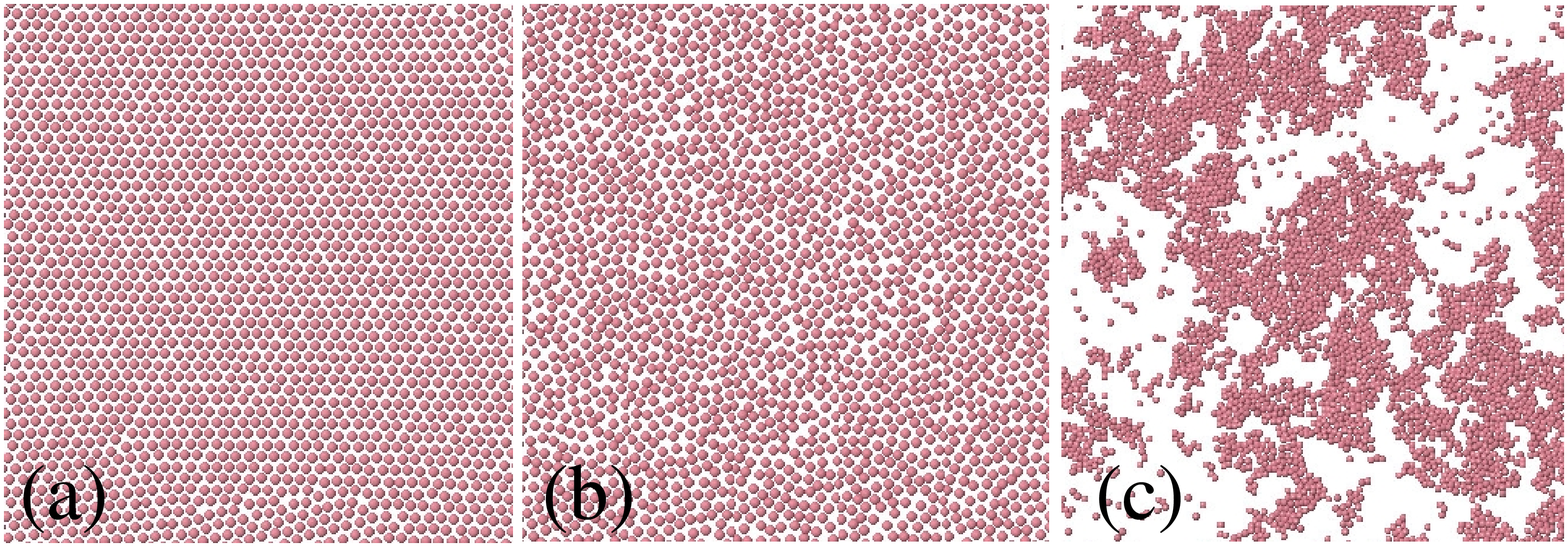,width=16.5cm}
\caption{\label{crystal} (a) A periodic crystalline structure 
does not flow because preserving the crystalline order
requires moving an extensive set of particles. 
(b) A mechanically rigid 
glassy structure exhibits neither the long-range order
of a crystal nor the large scale density fluctuations
observed at an ordinary critical point.
(c) Large scale critical density fluctuations near 
the critical point.}
\end{figure*}

From the point of view of statistical physics, glasses are
mysterious materials. Glassy materials possess a mechanical rigidity 
which is similar to the one of a crystalline material. 
In a crystal, rigidity is a direct consequence of 
the long-range periodic order: it is not possible to 
move a single particle in a perfect crystal (while preserving 
the crystalline order) without also moving an extensive set of neighbors,
see Fig.~\ref{crystal}(a). While mechanically rigid, 
glasses do not seem characterized by any type
of long-range order, see Fig.~\ref{crystal}(b), they actually 
resemble ordinary dense liquids. The comparison between crystals and glasses
suggests that perhaps a more subtle symmetry breaking takes place
during the formation of a glass, one that is not obvious to the
naked eye. This conundrum has been a long-standing issue 
in condensed matter physics~\cite{glassreview,rmp}.

Experimentally, one faces the fundamental difficulty that
liquids approaching the glass transition (for example by 
decreasing temperature) become too viscous to flow on
experimental timescales, and fall out of thermal equilibrium 
without encountering any reproducible thermodynamic phase 
transition. It is of course tempting to interpret this
dramatic dynamic slowing down as originating from an 
underlying phase transition or critical point. 
Near an ordinary critical point, large scale spatial 
fluctuations develop, as the density fluctuations in the example shown  
in Fig.~\ref{crystal}(c), and the dynamics slows down~\cite{dynamicrg}.
However, no such fluctuations are detected near the 
glass transition (see Fig.~\ref{crystal}(b)). Therefore, finding convincing
evidence of an underlying phase transition governing the physics
of amorphous materials would represent an important 
progress in this field.

In the last decade these questions have also come up in
the field of soft condensed matter, in which disordered
structures known as ``jammed'' materials~\cite{jammingbook}
(foams, emulsions, colloidal gels, sandpiles) 
stop flowing when their density becomes large, without possessing
long-range crystalline order, just like molecular glasses.
J.D. Bernal~\cite{bernal1} in the 60's was one of the first
physicists to suggest that disordered atomic fluids and granular
packings could be investigated using similar tools and, perhaps,
understood using similar theoretical concepts--an idea that
has since remained highly popular~\cite{liunagel}.

Two decades of research on dynamic heterogeneity (to be defined shortly)
in amorphous materials have established that the formation of rigid amorphous
structures is indeed accompanied by nontrivial spatio-temporal
fluctuations, which become stronger as the glassy
phase is approached and are characterized by growing dynamic
correlation length scales~\cite{book}. 
In this article we revisit the mounting
evidence -- using mostly the example of supercooled liquids,
where dynamic heterogeneity has been most widely analyzed
-- that the formation of amorphous materials is a complex
collective phenomenon, which shares more similarities with  
ordinary critical points than the featureless structure
shown in Fig.~\ref{crystal}(b) would suggest. 

\section{What is dynamic heterogeneity?}

The concept of dynamic heterogeneity as a key feature that
characterizes disordered materials has slowly emerged from
experimental studies of highly viscous molecular
liquids approaching the glass transition. In these systems,
relaxation spectra measured through mechanical or dielectric
probes span a very broad range of relaxation
times and are strongly nonexponential. This suggests the
existence of wide distributions of relaxation rates.

What is the microscopic origin of these broad distributions? 
Looking again at the disordered structure in 
Fig.~\ref{crystal}(b), it is natural
to imagine that the presence of structural disorder implies that 
atoms in different environments move differently.
The physical picture is that, at any given time, different regions in a liquid
might relax in a different manner and at a different rate, thus 
producing broad distributions of relaxation times. However, 
when the system is close to the glass transition but not yet a glass, 
particles constantly move and rearrange, and so the distinction 
between different spatial environments can only hold over a finite
duration. 

{\it Dynamic heterogeneity} refers to the existence of
transient spatial fluctuations in the local
dynamical behavior. 
Dynamic heterogeneity is 
observed in virtually all disordered systems with 
glassy dynamics~\cite{book}.

\begin{figure}[b]
\psfig{file=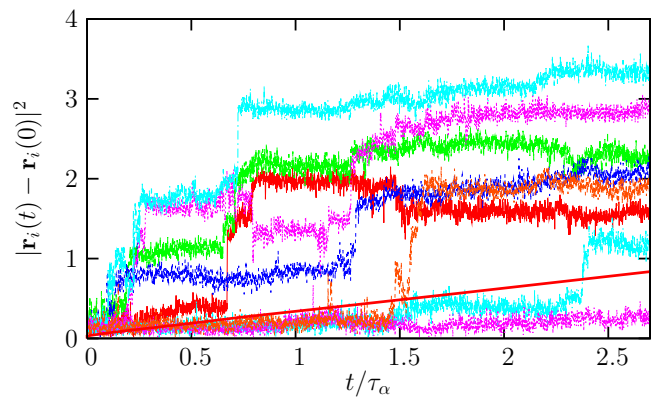,width=8.5cm}
\caption{\label{msd2} Time resolved squared displacements of individual 
particles in a simple model of a glass-forming liquid.
The average behavior is shown as the smooth full line. 
Individual trajectories are composed of long periods 
of time during which particles vibrate around well-defined 
positions, separated by rapid jumps that
are widely distributed in time and occur at different 
times and frequencies for different particles. }
\end{figure}

Direct confirmation of the heterogenous character of the dynamics stems
for instance 
from computer simulations of simple models of supercooled liquids, 
see Fig.~\ref{msd2}.
Whereas the averaged displacement of a particle in a given
time window of width $t$ is a smooth increasing function
of $t$, the time signals for individual particles shown
in Fig.~\ref{msd2} have two important 
characteristics. 

(i) They are highly intermittent in time, being composed
of a succession of long periods of time where particles
vibrate around well-defined locations, separated by
rapid ``jumps''. The waiting times separating successive jumps
are statistically broadly distributed.

(ii) The trajectories differ widely from one particle 
to another in the same system at the same time. Some particles undergo many 
jumps and move large distances while some other particles 
are nearly immobile over the entire time window.
 
As all other features related to dynamic heterogeneity,
such observations cannot be made from ensemble-averaged 
measurements. Indeed, the characterization of 
dynamic heterogeneity requires the development 
of experimental techniques that are not only sensitive 
to averaged or typical behaviour, but can also
resolve the {\it fluctuations}~\cite{book,ediger}. 
For theoreticians, the existence of dynamic heterogeneity
implies that fluctuations need to be taken into account 
in the description of transport properties. 
Therefore, materials close to a glass transition differ
qualitatively from ordinary fluids, where fluctuations 
can typically be neglected. 

The intermittency of single-particle trajectories, while a
clear indication of spatio-temporal fluctuations, do not
tell us how these fluctuations are correlated in space.
This point was first addressed in pioneering works using
four-dimensional NMR~\cite{nmr1} and direct probing
of fluctuations at the nanoscopic scale using atomic
force microscopy techniques in polymeric glasses~\cite{israeloff}.
Direct visualizations of molecular trajectories are not yet possible
but recent, very elegant, experimental approaches 
using single molecule spectroscopy are getting 
close to it~\cite{laura,bingemann}. Direct visualisation 
is possible for different types of glasses, such as colloidal~\cite{weeks}
and granular~\cite{dauchot,durian} assemblies.
All these spatially resolved
measurement indicate that extended regions of space indeed
transiently behave as fast and slow regions. 

\section{Need for high-order dynamic correlation functions}

\begin{figure}
\psfig{file=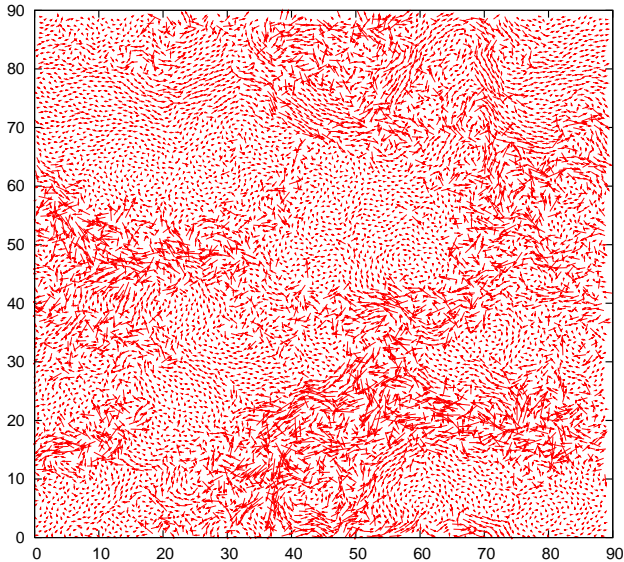,height=8.8cm,angle=-90}
\caption{\label{SHD} Spatial map of single particle displacements in 
the simulation of a Lennard-Jones model of a supercooled liquid 
in two spartial dimensions. Arrows represent the displacement
of each individual particle 
in a trajectory of duration comparable to the typical 
structural relaxation time. This map reveals that
particles with different mobilities are spatially correlated.} 
\end{figure}

We have described dynamic heterogeneity as spatial variations
in the local relaxation rate. Although hard to detect in
an experiment, these fluctuations are easily observed in
a computer simulation, where the position of
each particle is  known at each time step. In Fig.~\ref{SHD} we show an
example of the visualisation of the spatially heterogeneous
dynamics in a simple model of a supercooled liquid, in two spatial
dimensions. This
visualization shows the existence of spatially extended
``domains'' where the amplitudes of single-particle
displacements are correlated. Remember that 
these domains have no obvious counterpart in the 
density fluctuations, Fig.~\ref{crystal}(b), 
and only appear when {\it dynamics} is considered. 
However, the spatial fluctuations in 
Fig.~\ref{SHD} are obviously reminiscent of the critical
fluctuations in Fig.~\ref{crystal}(c), with one
major difference: while a thermodynamic quantity 
(the density) becomes critical in ordinary critical
phenomena, fluctuations are only detected through
dynamical quantities in highly viscous liquids. 

To characterize the spatial fluctuations shown 
in Fig.~\ref{SHD} in a statistical
manner, one must resolve dynamics in both space and time 
and quantify deviations from the average behavior.
To this end, one defines a ``mobility'', $c_i(t,0)$, which 
quantifies how much particle $i$ moves between times 
$t=0$ and $t$ [for instance $c_i(t) = e^{-|{\bf r}_i(t)-{\bf r}_i(0)|^2 
/  d^2}$, with $d$ of the order of the particle size].
Given two particles at separation $r$, one can measure the
degree to which their mobilities are correlated.  
It is convenient to define a ``mobility field'' for a system 
composed of $N$ particles via
\be
c({\bf r} ; t,0) = \sum_{i=1}^N c_i(t,0) \delta[{\bf r}-{\bf r}_i(0)],
\ee  
where ${\bf r}_i(t)$ is the position of particle $i$ at time $t$.
The spatial
correlations of the mobility are finally captured by
the correlation function
\be
G_4(r;t) = \langle c({\bf r};t,0) c({\bf 0};t,0) 
\rangle - \langle c({\bf r};t,0) \rangle^2,
\label{equ:g4def}
\ee
which depends only on the time $t$ and the distance
$r = |{\bf r}|$ as long as the ensemble 
average, denoted with brackets, 
is taken at equilibrium in a translationally invariant
system; $G_4(r;t)$
is known as a `four-point' dynamic correlation function
because it measures correlations of motion between time 0 and time $t$
arising at two points, ${\bf 0}$ and ${\bf r}$. 

The analogy with fluctuations in critical systems 
becomes clear in Eq.~(\ref{equ:g4def}) if one considers
the mobility field $c({\bf r};t,0)$ as playing the role of the 
order parameter for the transition, characterised by 
nontrivial fluctuations and correlations near the glass 
transition. This analogy is now fully exploited
in modern theoretical treatments~\cite{rmp}. 

The above definition of a real-space correlation function of the mobility
represents a vital advance in the characterization of dynamical
heterogeneity. For instance, it allows the language of field
theory and critical phenomena to be used in studying
dynamical fluctuations in glassy systems~\cite{steve,BBEPL}. 
Like in critical phenomena, if one makes the hypothesis 
that there exists a single dominant length scale
$\xi_4$ then one expects that for large distances the correlation
function $G_4(r;t)$ decays as 
\be 
G_4(r; t) \approx \frac{A(t)}{r^{p}} e^{-r/\xi_4(t)},
\label{eqscaling}
\ee
with $p$ a critical exponent. 
It is also natural to define the susceptibility associated with
the correlation function
\be
\chi_4(t) = \int\mathrm{d}^d {\bf r}\, G_4(r;t).
\ee
If the prefactor $A(t)$ in Eq.~(\ref{eqscaling})
is known, the susceptibility $\chi_4(t)$
can be used to infer the typical number of particles
involved in the correlated motion shown in Fig.~\ref{SHD}. 
That is, $\chi_4(t)$ may
be interpreted as the ``volume'' of the correlated clusters.

Further, $\chi_4(t)$
can also be obtained from the fluctuations of 
the total mobility $C(t,0)=\int\mathrm{d}^d{\bf r}\, c({\bf r};t,0)$, 
through
\be
\chi_4(t) = N [ \langle C(t,0)^2 \rangle - \langle C(t,0) \rangle^2 ].
\label{equ:chi4var}
\ee
In practice, this formula allows an efficient measure of
the degree of dynamical heterogeneity, at least in computer
simulations and in those experiments where the dynamics can be
spatially and temporally resolved. As long as $c({\bf r};t,0)$
appropriately quantifies atomic motion, 
$\chi_4(t)$ can be measured in a 
variety of systems, serving as a basis for fair comparisons
of the extent of dynamical heterogeneity and has 
become a central tool in characterizing dynamic heterogeneity
in amorphous materials~\cite{book}.

\section{Four-point susceptibilities in molecular,
colloidal, and granular glasses}

\begin{figure}
\begin{center}
\psfig{file=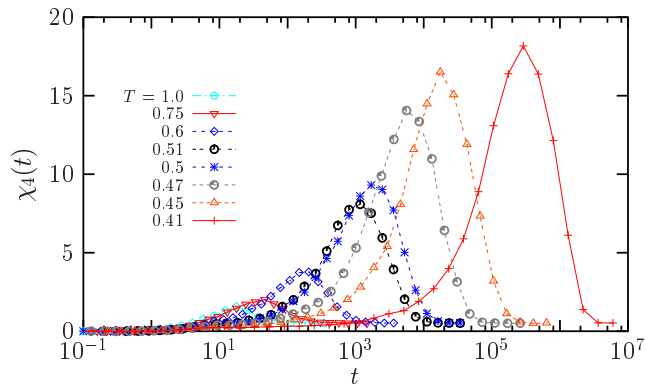,width=8.5cm}
\end{center}
\caption{\label{chi4ludo} Time dependence of the susceptibility
$\chi_4(t)$ that quantifies spontaneous fluctuations of the
relaxation function in a molecular dynamics
simulation of a supercooled liquid.  For each temperature,
$\chi_4(t)$ has a maximum, which shifts to larger times and has
a larger value when $T$ is decreased, revealing the increasing
length scale of dynamic heterogeneity in supercooled liquids
that approach the glass transition.} 
\end{figure}

The dynamical function $\chi_4(t)$ has now 
been measured in computer simulations
of many different glass-forming liquids, by molecular dynamics,
Brownian, and Monte Carlo 
simulations~\cite{dasgupta-chi4,yamamoto,glotzerfranzparisi,lacevic,glotzer,berthier}. 
An example is shown in Fig.~\ref{chi4ludo} for a Lennard-Jones 
numerical model, but
the qualitative behavior is similar in all 
cases~\cite{franzparisi,TWBBB,jcpI}: 
as a function of time $\chi_4(t)$ 
increases at first, it has a peak on a timescale of the order of the typical
relaxation time of the fluid,  
and then it decreases at large times. This time dependence 
simply reflects the 
the transient nature of the dynamical heterogeneity.

The peak value of $\chi_4(t)$ approximately measures the volume 
over which structural relaxation processes are correlated.
Therefore, the most important result obtained from data
such as those presented in Fig.~\ref{chi4ludo} is the temperature 
evolution of the peak height, which is found to increase when 
the temperature decreases and the global dynamics slows down.
Such data provides direct evidence that the 
approach to the glass transition is accompanied by the 
development of increasingly long-ranged spatial
correlations of the dynamics.  

In experiments, direct 
measurements of $\chi_4(t)$ have been made in
colloidal~\cite{weeks2} and granular materials~\cite{dauchot,durian} 
close to the colloidal and granular glass transitions,
and also in foams~\cite{mayer} and gels~\cite{luca}, 
because dynamics is more easily spatially 
and temporally resolved in those cases.
The results obtained in all these cases are again broadly similar 
to those shown in Fig.~\ref{chi4ludo}, both for the 
time dependence of $\chi_4(t)$ and its evolution with
a change of the relevant variable controlling the dynamics. 

Obtaining information on the behavior of $\chi_4(t)$
and $G_4(r;t)$ from experiments on molecular systems
is difficult, because it is hard to disentangle the spontaneous 
fluctuations embodied in $\chi_4(t)$ in Eq.~(\ref{equ:chi4var})
from the experimental noise. 
Such measurements are however crucial because numerical
simulations and experiments on colloidal and granular systems
can typically only be performed for relaxation times spanning
at most 5-6 decades.  On the other hand, in molecular liquids,
up to 14 decades are in principle relevant, and extrapolation of
simulation data all the way to the experimental glass transition
is fraught with difficulty.  Indirect estimates of $\chi_4(t)$
from experiments are discussed below.

\section{Real-space measurements and dynamic structure factors}

We mentioned above that a growing peak in $\chi_4(t)$ ``directly''
reveals the growth of a dynamic correlation length scale
as the glass transition is approached. This can only be true 
if the assumptions made in Eq.~(\ref{eqscaling}) for the scaling form of
$G_4(r;t)$ are correct. Dynamic length scales
should in principle be obtained by direct
measurements of a spatial correlation function.   
 
However, in contrast to $\chi_4(t)$, detailed measurements of $G_4(r;t)$
are technically more challenging as dynamic correlations
must now be resolved in space over a large range of distances with a very high
precision, and so there is much less data to draw on.
From the point of view of numerical simulations where 
many measurements of $\chi_4$ were reported, the main limitation 
to properly measuring $\xi_4$ is the system size. This might seem surprising 
as typical numbers extracted for the correlation
length scale $\xi_4$ are rather modest, but a precise 
determination of $\xi_4$ requires an accurate study of the tail of 
$G_4(r;t)$ at large $r$, which entails an important 
numerical effort~\cite{sri,szamel}.

However, such studies are important in that they allow
the dynamical length scale $\xi_4(t)$ to be measured directly.
Moreover, such studies help infer 
the behavior of $\xi_4(t)$ from measurements of $\chi_4(t)$.
Published work is consistent with $\chi_4(t)/G_4(0,t)$
representing the number of particles involved in heterogeneous
relaxation.  Therefore, truly `direct' measurements indeed
confirm that the increase of the peak of $\chi_4(t)$
corresponds, as expected, to a growing dynamic length scale
$\xi_4(t)$~\cite{glotzer,berthier,jcpI,sri,szamel}.

Instead of direct inspection of $G_4(r;t)$, it is often
convenient to analyse its Fourier transform,
\be
S_4(q;t) = \int\mathrm{d}^d {\bf r} e^{i {\bf q} \cdot {\bf r}} G_4(r;t).
\ee
which is known as the four-point
structure factor of dynamic heterogeneity.
In Fourier space, the 
large domains observed in Fig.~\ref{SHD} impact the low
wavevector behavior of $S_4(q;t)$ 
in the form of a peak that grows when the glass
transition is approached.  This peak is often fitted 
with the Ornstein-Zernike functional form which is frequently used in 
conventional critical phenomena~\cite{dynamicrg}.

\section{Experimental estimates of multipoint susceptibilities}
\label{inequality}

Although readily accessible in numerical simulations, the 
fluctuations that give access to $\chi_4(t)$ 
are in general very small and impossible 
to measure directly in 
experiments, except when the range of the dynamic correlation is 
macroscopic, as in granular materials~\cite{dauchot} or in
soft glassy materials where 
it can reach the micrometer and 
even millimetre range~\cite{mayer,luca,duri09}.
To access $\chi_4(t)$ in molecular liquids, one should
perform time-resolved dynamic measurements probing very small volumes,
with a linear size of the order of a few nanometers.

Fortunately, simpler alternative procedures exist. The central idea
underpinning these solutions is the realizations that 
if it is generally hard to detect noise in an experiment, 
it is usually simpler to measure the response of a system to 
an external perturbation. In the linear response 
regime, both types of measurements can often be related to one another
by fluctuation-dissipation theorems~\cite{dynamicrg}. The physical motivation
is that while four-point correlations offer a direct probe
of the dynamic heterogeneities, other multi-point correlation
functions might also give useful information about the microscopic
mechanisms leading to these heterogeneities.

For example, one might expect that a local fluctuation of the
enthalpy $\delta h_x(t=0)$ at position $x$ 
and time $t=0$ triggers or eases the dynamics in its surroundings, 
leading to a systematic correlation between 
$\delta h_x(t=0)$ and $c(x+r;t,0)$. A similar effect 
is expected for a local fluctuation of the density. These 
physical intuitions suggest the definition of a family of 
three-point correlation functions that relate thermodynamic 
fluctuations at one point to dynamical ones at another point. 
Crucially, some of these 
three-point correlations are both experimentally accessible and give 
bounds or approximations to the four-point dynamic 
correlations~\cite{science}. 

Based on this insight, one may obtain a lower bound 
on $\chi_4(t)$
by using a fluctuation-dissipation relation, which is valid 
at equilibrium when the energy
is conserved by the dynamics~\cite{science}: 
\begin{equation}
\label{eq7:equation}
\chi_{4}(t) \geq \frac{k_{\rm B}T^{2}}{c_{P}} \left[ \chi_{T}(t) 
\right]^{2},
\end{equation}
where $\chi_T(t)$ quantifies the {\it response} of
the {\it average} mobility to an infinitesimal change in the 
temperature $T$, and $c_P$
is the specific heat per particle.
The response $\chi_{T}(t)$ can be experimentally accessed
by measuring the average of a dynamical correlator, 
$\langle C(t,0) \rangle_T$, 
at nearby temperatures, $T$ and $T + \delta T$, in the linear 
regime $\delta T \ll T$:
\be
\chi_T(t) \approx \frac {\langle C(t,0) \rangle_{T+\delta T} -  \langle 
C(t,0) \rangle_{T}}{\delta T}.
\label{chit}
\ee 
The main experimental advantage of (\ref{chit}) is that 
spatio-temporal resolution is not needed, contrary 
to Eq.~(\ref{equ:chi4var}). 

Detailed numerical simulations and theoretical
arguments~\cite{jcpI} strongly suggest that the right
hand side of Eq.~(\ref{eq7:equation}) actually provides a good
estimate of $\chi_4(t)$ in supercooled liquids, and not just
a lower bound.  Similar estimates exist considering density
instead of the temperature in Eq.~(\ref{chit}). These are
useful when considering colloidal or granular materials where
the glass transition is mostly controlled by the packing
fraction. The quality of the corresponding lower bound was
tested experimentally on granular packings close to the jamming
transition \cite{lech2}, and numerically for colloidal hard
spheres~\cite{gio,szamel}.

\begin{figure}
\begin{center}
\psfig{file=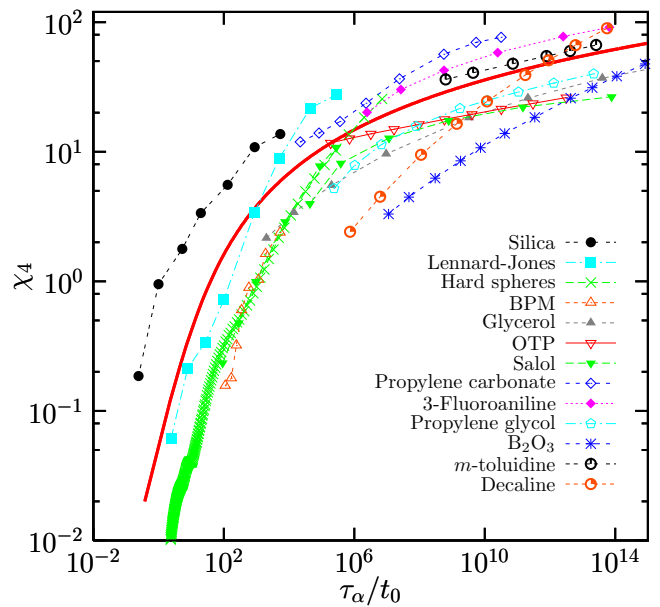,width=8.5cm}
\end{center}
\caption{\label{cecile} 
``Dynamic scaling'' relation 
between the number of dynamically 
correlated particles, evaluated by the peak height of $\chi_4$, 
and relaxation timescale, $\tau_\alpha$, for a number of 
glass-formers~\cite{cecile}, determined using the 
bound provided by Eq.~(\ref{eq7:equation}). For all systems, dynamic 
correlations increase when the glass transition is 
approached. The full line through the data~\cite{cecile} 
suggests a crossover from 
algebraic, $\chi_4 \sim \tau_\alpha^z$,
to logarithmic, $\chi_4 \sim \exp (\tau_\alpha^\psi)$,
growth of dynamic correlations with increasing $\tau_\alpha$.} 
\end{figure}

We show in Fig.~\ref{cecile} a compilation of data~\cite{cecile}
for the evolution of the peak height of $\chi_4$, in a 
representation inspired by the theory of dynamic 
critical phenomena~\cite{dynamicrg}.  These data 
represent an experimental confirmation that 
dynamic fluctuations and correlation length scales grow appreciably
when molecular liquids approach the glass transition. However, we also
learn that typical length scales do not become very large 
(remaining in the nanometer scale for molecular glass-formers)
before liquids vitrify in a 
nonergodic state, and that an ``ideal'' glass critical
point is not readily accessible to experiments. 

\section{Theoretical developments}

The above results are also relevant because many theories of
the glass transition have assumed or predicted that the dynamics
slows down because there are increasingly large regions over
which particles have to relax in a correlated 
manner~\cite{glassreview}. However, in the
absence of experimental signs of growing length scales, these
theoretical constructions would have remained speculative.

The measurements of the spatial extent of dynamic heterogeneity, 
in particular $\chi_4(t)$ and  $G_4(r;t)$, seem  
to provide the long-sought evidence that the glass transition must indeed 
be considered as a form of critical or collective 
phenomenon involving growing timescales and lengthscales.
This is important for the field of
glass transition, even though a conclusive understanding of
the relationship between dynamical length scales and relaxation
timescales is still the focus of intense research~\cite{rmp}.

From a theoretical perspective, we are
familiar with the idea, borrowed from equilibrium
critical phenomena~\cite{dynamicrg}, 
that when correlation length scales get large,
microscopic features of the system become unimportant and 
``universal'' behaviors emerge. Whether realistic glassy systems
have length scales that are large enough for such a universal
description remains unclear.  Although most theoretical 
approaches are in this spirit, one should perform
an equally careful treatment of pre-asymptotic effects, which 
obviously matter for experiments that are performed far 
away from (putative) criticality.
Therefore, theories of the glass transition are still crude 
descriptions of reality, despite large research efforts.

Distinct microscopic mechanisms have been proposed that all
give rise to growing dynamic correlations similar to the ones
revealed by four-point functions. This relative abundance of
approaches can be interpreted in two ways. A first view is to consider 
dynamic heterogeneity as a unifying physical phenomenon 
that accompanies the formation of glasses, and is central 
to its theoretical understanding.
The second, more pessimistic interpretation, is that 
if any reasonable theory predicts increasingly
spatially heterogeneous dynamics, then studying 
dynamic heterogeneity might not help make substantial
theoretical progress. This argument is balanced by
the fact that the mere observation of dynamic heterogeneity
has forced theoreticians to work hard on many fronts to
understand what sort of microsopic mechanism is able (or not)
to provide for instance new insights or quantitative predictions
for the behavior of multipoint correlation functions,
susceptibilities and dynamic correlation length scales~\cite{book}. 

Several
theoretical schemes, such as mode-coupling theory~\cite{mct}, 
the random first-order transition theory~\cite{rfot}, 
dynamic facilitation approach~\cite{kcm},
and frustration-limited domain scaling picture~\cite{gilles} have now been
developed to the point that they provide useful theoretical
guides to understand and analyze various aspects of dynamic
heterogeneity, see \cite{rmp} for a general theoretical review.  
One can thus hope that these approaches contain
useful seeds for the construction of a ``unified'' theory of
the glassy state. In all of these approches, dynamic heterogeneity 
and spatio-temporal fluctuations feature as central concepts. 

\section{Perspectives}

Although we mostly discussed 
supercooled liquids near the molecular glass transition, we must 
emphasize that other materials have played an important
role in developing the concepts and tools described above. 
For instance, several soft condensed materials such as colloidal
assemblies have been instrumental in understanding the 
phenomenon of dynamic heterogeneity, because they 
share important similarities with molecular liquids but their
elementary constituents are so much larger than atoms 
(in the range 50 nm -- 1$\mu$m) that they can more easily 
be visualized~\cite{weeks2,weeks}. 
Driven granular media (sheared or agitated systems) also
undergo a ``granular'' glass transition which is empirically 
similar to the molecular transition. In that case, grains can
even be directly tracked using a standard 
camera~\cite{dauchot,durian}. These studies
have contributed to give flesh to the concept of dynamic 
heterogeneity, because spatial fluctuations could be {\it seen}.

In the introduction we alluded to 
the rigidity transition occurring in athermal 
disordered granular packings.
This ``jamming'' transition arises in an assembly of rigid particles 
when the system cannot be compressed anymore, and is thus 
mainly a geometric transition where thermal 
fluctuations play no role.
For spherical particles of equal sizes, this transition 
occurs near `random close packing' 
$\varphi_{\rm rcp} \approx 0.64$~\cite{bernal1}.
The jamming transition is relevant also 
for athermal assemblies of soft particles, such as foams and 
emulsions, which are thus additional examples 
of disordered rigid materials~\cite{jammingbook}.  
Connections with the physics of 
glasses are still rather speculative 
but are currently the focus on an important research effort.
Detailed studies of dynamic
heterogeneity in packings of soft and hard particles
near random close packing started to appear only 
recently~\cite{lechenault2,trappe4,claus,durianarxiv}, 
and could help elucidating similarities and differences
between glasses and granular materials. 

One of the most frequently asked questions in studies of dynamical
heterogeneity is whether the observed dynamic fluctuations and 
correlations might have a structural origin: 
Is there, after all, a ``hidden'' thermodynamic
order parameter which would exhibit spatial fluctuations 
comparable to the ones revealed by dynamic heterogeneity 
studies?  This question has attracted sustained
interest. For example, in very early numerical 
work on dynamic heterogeneity, immobile regions were discussed 
in connection with compositional fluctuations in fluid 
mixtures~\cite{hurley}. However, dynamic heterogeneity
would not have emerged as an important concept 
if a simple, direct connection between 
structural order and relaxation dynamics 
had been satisfactorily established in amorphous materials.
In that case, research would be dedicated to 
understanding the development of structural correlations
at low temperatures in supercooled liquids, and to developing 
tools to measure, quantify and analyse such static 
features. 

Having said this, recent research on 
isoconfigurational ensembles~\cite{iso}, 
amorphous order~\cite{jorge} 
and point-to-set correlations~\cite{bb}, 
suggests that the structure of disordered materials 
might well be the next topic where new 
discoveries and concepts will emerge in the near future.
While two-point static correlations
are poorly correlated to the evolution of the glassy 
dynamics, there is plenty of room for inventing more complicated 
correlation functions that could more accurately characterize 
the local structure of complex disordered media and explain 
their fascinating physical properties.

\end{document}